\documentclass[twocolumn]{emulateapj}

\usepackage{amsmath}%
\usepackage{amssymb}%
\usepackage{srcltx}
\usepackage[english]{babel}
\usepackage{hyperref}
\usepackage{graphicx}

\newcommand{\noprint}[1]{}

\begin{document}

\selectlanguage{english}

\title{The Outburst of the Blazar \object{S4~0954+658} in March-April 2011}
\author{D.A.\,Morozova\altaffilmark{1}, V.M.\,Larionov\altaffilmark{1,2}, I.S.\,Troitsky\altaffilmark{1}, 
S.G.\,Jorstad\altaffilmark{1,3},  A.P.\,Marscher\altaffilmark{3},  J. L.\,G\'{o}mez\altaffilmark{9}, 
 D.A.\,Blinov\altaffilmark{4,1}, N.V. \,Efimova\altaffilmark{1,5}, V.A.\,Hagen-Thorn\altaffilmark{1,2}, E.I.\,Hagen-Thorn\altaffilmark{1,5}, M.\,Joshi\altaffilmark{3}, T.S.\,Konstantinova\altaffilmark{1}, E.N.\,Kopatskaya\altaffilmark{1},  L.V.\,Larionova\altaffilmark{1}, E.G.\,Larionova\altaffilmark{1},  A.\,L\"{a}hteenm\"{a}ki\altaffilmark{6,7}, J.\,Tammi\altaffilmark{6}, E.\,Rastorgueva-Foi\altaffilmark{6}, I.\,McHardy\altaffilmark{8}, M.\,Tornikoski\altaffilmark{6}, I.\,Agudo\altaffilmark{10,9,3}, C.\,Casadio\altaffilmark{9}, S.N.\,Molina\altaffilmark{9}, A. E.\,Volvach\altaffilmark{11}, L. N.\,Volvach\altaffilmark{11} 
}

\altaffiltext{1}{Astronomical Institute of St. Petersburg State University, Universitetsky Pr. 28, Petrodvorets, 198504, St. Petersburg, Russia; comitcont@gmail.com}
\altaffiltext{2}{Isaac Newton Institute of Chile, St. Petersburg Branch}
\altaffiltext{3}{Institute for Astrophysical Research, Boston University,
725 Commonwealth Ave., Boston, MA 02215-1401; jorstad@bu.edu}
\altaffiltext{4}{Department of Physics , University of Crete, 71003, Heraklion, Greece} 
\altaffiltext{5}{Pulkovo Observatory, Russian Academy of Sciences, Pulkovskoe sh. 65, St. Petersburg, 196140 Russia}
\altaffiltext{6}{Aalto University Mets\"ahovi Radio Observatory, Mets\"ahovintie 114,FIN-02540 Kylm\"al\"a, Finland.}
\altaffiltext{7}{Aalto University Dept of Radio Science and Engineering, PL 13000,
FIN-00076 Aalto, Finland}
\altaffiltext{8}{Department of Physics and Astronomy, University of Southampton, Southampton, SO17 1BJ,
United Kingdom}
\altaffiltext{9}{Instituto de Astrof\'{i}sica de Andaluc\'{i}a, CSIC, Apartado 3004, 18080, Granada, Spain}
\altaffiltext{10}{Joint Institute for VLBI in Europe, Postbus 2, NL-7990 AA Dwingeloo, the Netherlands} 
\altaffiltext{11}{Radio Astronomy Laboratory of the Crimean Astrophysical Observatory, Katsiveli, Crimea, 98688 Ukraine}

\shorttitle{The Outburst of the Blazar S4~0954+658 in March-April 2011}
\shortauthors{Morozova et al.}
\begin{abstract}
We present the results of optical ($R$ band) photometric and polarimetric monitoring and Very Long Baseline Array (VLBA) imaging of the blazar \object{S4~0954+658}, along with \textit{Fermi} $\gamma$-ray data during a multi-waveband outburst in 2011 March-April. After a faint state with a brightness level R $\sim$17.6 mag registered in the first half of January 2011, the optical brightness of the source started to rise and reached $\sim$14.8 mag during the middle of March, showing flare-like behavior. The most spectacular case of intranight variability was observed during the night of 2011 March 9, when the blazar brightened by $\sim$0.7 mag within ~ 7 hours. During the rise of the flux the position angle of optical polarization rotated smoothly over more than 300\degr. At the same time, within $1~\sigma$ uncertainty a new superluminal knot appeared with an apparent speed of $19.0\pm 0.3~c$. We have very strong evidence for association of this knot with the multi-waveband outburst in 2011 March-April.
We also analyze the multi-frequency behavior of  \object{S4~0954+658} during a number of minor outbursts from August 2008 to April 2012.  We find some evidence of connections between at least two more superluminal ejecta and near-simultaneous optical flares.

\end{abstract}

\section{Introduction}
The blazar S4\,0954+658 (z=0.367) is a well studied BL Lac object at optical wavelengths.
Its optical variability was analyzed by \citet{1993A&A...271..344W}, who found 
large amplitude variations (of $\sim 100$\%) on time scale as short as 1 day. 
\citet{1999A&A...352...19R} presented a comprehensive study of the optical and 
radio variability of the source during 1994-1998. They detected large amplitude intranight 
variations. An investigation of $B-R$ color variations allowed them to conclude that 
mid- and long-term brightness variations of the source are not associated with spectral variability.
\citet[and references therein]{2000MNRAS.315..229G} analyzed the radio morphology 
of S4\,0954+658  and showed that the jet is bent on both parsec and kiloparsec jet scales. 
They also found substantial intranight polarization variability of the radio core at 5~GHz. \citet{2010evn..confE..45K} have found several moving components in the jet at 22 GHz with mean velocity $4.9\pm0.4$~c. However, the kinematics of the parsec-scale jet of S4\,0954+658 is poorly studied, especially at 43 GHz. 

According to \citet{1995ApJ...445..189M}, $\gamma$-ray emission of S4\,0954+658 first was detected by EGRET in 1993. S4\,0954+658 was also detected by Fermi LAT according to the Fermi first and second catalogs of $\gamma$-ray bright sources \citep{2010ApJS..188..405A,2012ApJS..199...31N}.
In this paper we present a detailed study of the optical outburst of S4\,0954+658
in 2011 March-April \citep{2011ATel.3220....1L} along with an analysis of the $\gamma$-ray variability and 
behavior of the innermost radio jet at 43~GHz. Preliminary results of our observations 
have been described by \citet{2011arXiv1110.5861L}.

\section{Observations and Data Reduction}\label{obs}
The observations reported here were collected as a part of a long-term multi-wavelength 
study of a sample of $\gamma$-ray bright blazars. An overview of this program 
is given by \citet{2012IJMPS...8..151M}.

\subsection{Optical Observations}
We carry out optical $BVRI$ observations at the 70-cm AZT-8 reflector of the Crimean Astrophysical 
Observatory, and  40-cm LX-200 telescope in St.~Petersburg, Russia. The telescopes are equipped 
with identical photometers-polarimeters based on ST-7 CCDs. We perform observations in 
photometric and polarimetric modes at the 1.8-m Perkins telescope of Lowell Observatory (Flagstaff, AZ) using the PRISM camera
and at the 2.2~m telescope of Calar Alto Observatory (Almer\'ia, Spain) within the MAPCAT program\footnote{
\url{http://www.iaa.es/~iagudo/research/MAPCAT/MAPCAT.html}}. 
Photometric measurements in $R$ band are supplemented by observations at 
the 2-m Liverpool Telescope at La Palma, Canary Islands, Spain. 
Polarimetric observations  at the AZT-8, Perkins, and  Calar Alto telescopes are carried out in Cousins $R$ band,
while at the LX-200 telescope they are performed in white light, with effective wavelength close to $R$ band.

The Galactic latitude of S4\,0954+658 is 43\degr and  $A_V=0\fm38$, so that the interstellar polarization (ISP) in this direction is less 
than 1\%. To correct for the ISP, the mean relative Stokes parameters of nearby stars 
were subtracted from the relative Stokes parameters of the object.  This accounts for the instrumental 
polarization as well, under the assumption that the radiation of the stars is unpolarized. 
The errors in the degree of polarization, $P$, are less than 1\% (in most cases less than 0.5\%), 
while the electric vector position angle (EVPA), $\chi$, is determined with an uncertainty of 1--2\degr. 
The photometric errors do not exceed $0\fm02$.
Photometry and polarimetry of the source during the flare are presented in Table \ref{tbl-1}. 

\subsection{Gamma-ray Observations}
We derive $\gamma$-ray flux densities at 0.1-200 GeV by analyzing data from the Fermi Large Area Telescope (LAT), provided 
by the Fermi Science Space Center using the standard software \citep{2009ApJ...697.1071A}. We have constructed 
$\gamma$-ray light curves with a binning size of 7 days, with a detection criterion that the maximum-likelihood 
test statistic (TS) should exceed 10.0. Although the $\gamma$-ray flux fell below the detection limit during most 
of the period of our observations ($\le 5\times10^{-7}\mbox{ph}\cdot \mbox{cm}^{-2}\mbox{s}^{-1}$ ), 
there are a number of positive $\gamma$-ray detections  that are interesting to compare with behavior of the source at other wavelengths.

\subsection{Single-Dish Radio Observations}

We use 37 GHz observations obtained with 13.7~m telescope at Mets\"ahovi Radio Observatory of Aalto University, Finland. The flux density calibration is based on observations of DR~21, with 3C~84 and 3C~274 used as secondary calibrators. A detailed description of the data reduction and analysis is given in \citet{1998A&AS..132..305T}. These data are supplemented by observations carried out at the 22-meter RT-22 radio telescope of the Crimean Astrophysical Observatory at 36.8 GHz.  In this case the sources 2037+421, 1228+126,  and 2105+420 are used for the flux density calibration.  A detailed description of the data reduction and analysis can be found in \citet{2000AstL...26..204N}. 

\subsection{VLBA Observations}
The BL Lac object S4\,0954+658 is monitored monthly by the BU group with the VLBA 
at 43~GHz within a sample of bright $\gamma$-ray blazars\footnote{\url{http://www.bu.edu/blazars}}. 
The VLBA data are calibrated and imaged in the same manner as discussed in \citet{2005AJ....130.1418J}.
We have constructed total and polarized images at 33 epochs from 2010 August to 2012 April.
Each image in Stokes $I, Q,$ and $U$ parameters was fit by a model consisting of a number of components with 
circular Gaussian brightness distributions. Identification of components in the jet across epochs is based on 
analysis of their flux, position angle, distance from the core, size, degree of polarization, and EVPA. 
During this period we have identified 12 components, A1, K1, K2, K3, K4, K5, K6, K7, K8, K9, K10, K11, in addition 
to the core, A0. The core is a stationary feature located at the southern end of the portion of the jet 
that is visible at 43 GHz. We have computed kinematic parameters of knots (the proper motion, velocity, acceleration) by fitting the (x, y) positions of a component over epochs by different polynomials of order from 1 to 3, in the same manner as described in \citet{2005AJ....130.1418J}. The method produces uncertanties of polynomial coefficients with an assumption that the true value lies with probability $W$ within the confidence region around the estimated value ($W$=0.95 is applied). The ejection time of a component is the extrapolated time of coincidence of the position of a moving knot with the core in the VLBA images, and $T_{\mathrm{eject}}$ is the average of $T_{\mathrm{xeject}}$ and $T_{\mathrm{yeject}}$ weighted by their uncertainties, which are calculated using uncertainties of the polynomial coefficients.

Table~\ref{tbl-3} lists for the core and each superluminal knot the flux, fractional polarization level, $p$, and EVPA, $\chi$. Table~\ref{tbl-4} lists for each superluminal knot the apparent speed, $\beta_{app}$, acceleration, if detected ($\dot{\mu_{||}}$ and $\dot{\mu_\perp}$, along and perpendicular to the jet, respectively), 
mean position angle with respect to the core, $<\Theta>$, and extrapolated time of zero separation from the core, $T_{\mathrm{eject}}$. 

\section{Results and discussion}
\subsection{Optical Polarization Analysis}
\begin{figure}[ht]
\begin{center}
   \includegraphics[width=8cm,clip]{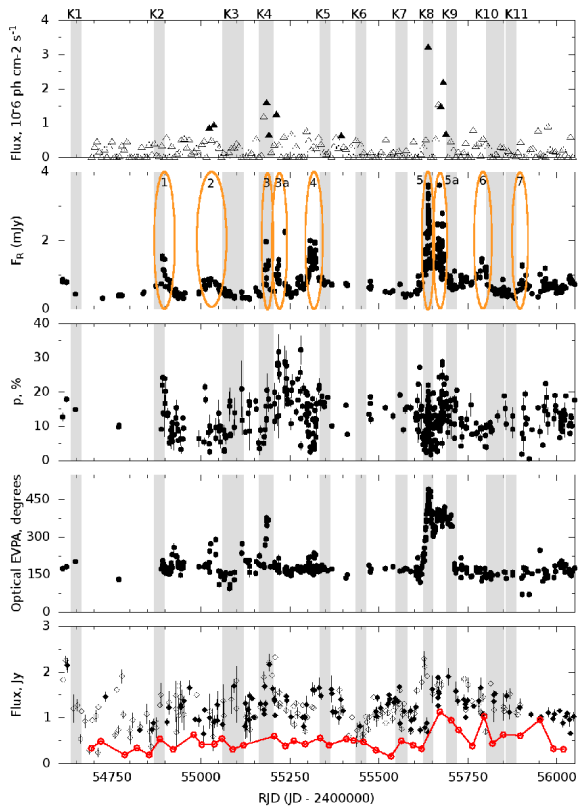}
        \caption{From top to bottom: $\gamma$-ray light curve (open triangles are the upper limits); optical (R band) light curve; fractional polarization vs. time; 
        position angle of polarization vs time; light curve of the VLBI core 
        at 43 GHz (open circles), and light curve from the whole source at 37~GHz (filled diamonds) and 35~GHz (open diamonds). The vertical
        bars show the times of ejection of superluminal knots within 1-$\sigma$ uncertainty. \label{fig1}}
\end{center} 
\end{figure}

\begin{figure}[ht]
\begin{center}
   \includegraphics[width=8cm,clip]{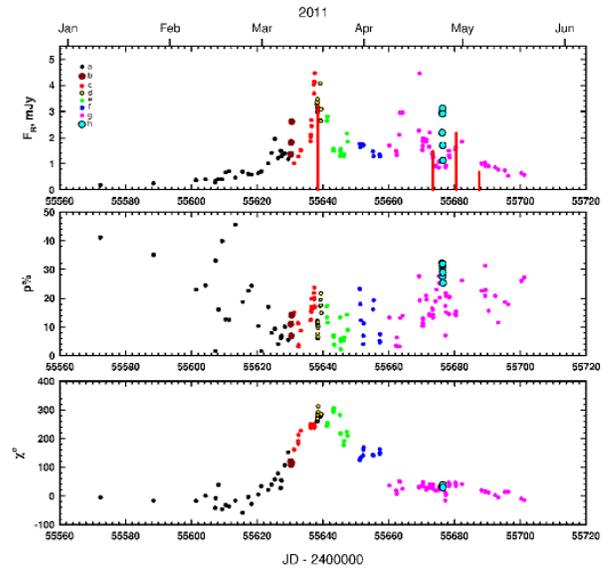}
        \caption{Optical flux density (corrected for Galactic extinction), fractional polarization, and position angle of polarization
        in R band vs. time in 2011 January-May; magnified symbols refer to the nights with violent 
        intranight variability, colors designate sections of the data with different Stokes parameter behavior (see Table \ref{tbl-2}). Red vertical bars in the upper panel mark positive Fermi LAT detections, a bar's height is proportional to the $\gamma$-ray flux.\label{fig2}}
\end{center} 
\end{figure}

Figure~\ref{fig1} displays the entire set of optical photometric and polarimetric data collected by our team 
during 2008-2011.
The blazar shows prominent activity during the period covered by our observations, 
with the $R$ band amplitude of variations exceeding $2^{\rm m}$  and a record level of $P$  exceeding 40\%. 
Even on such an active background the outburst, which started in early 2011, 
is quite prominent. An enlargement of the event is shown in Figure~\ref{fig2}. 

\begin{deluxetable*}{cccccccc}

\tablecaption{Photometry and polarimetry of S4~0954+658 during 2011 April-May outburst \label{tbl-1}}

\tablehead{\colhead{RJD} & \colhead{$R$} & \colhead{$\sigma R$} & \colhead{$p$} & \colhead{$\sigma p$} & \colhead{$EVPA$} & \colhead{$\sigma EVPA$} & \colhead{Telescope} \\ 
\colhead{(days)} & \colhead{(mag)} & \colhead{(mag)} & \colhead{(\%)} & \colhead{(\%)} & \colhead{($\degr$)} & \colhead{($\degr$)} & \colhead{} } 

\startdata
 55577.4520 & 17.105 & 0.014 & 15.57 &  0.44 & 172.5 &   0.1 & CAHA\\
 55588.5160 & 17.182 & 0.016 & 15.48 &  1.12 & 163.1 &   2.1 & AZT-8+ST7\\
 55601.4340 & 16.924 & 0.010 & 12.55 &  0.70 & 162.2 &   1.6 & AZT-8+ST7\\
 55603.9170 & 16.985 & 0.007 & 13.54 &  0.57 & 148.3 &   1.2 & Perkins\\
 55604.3090 & 16.896 & 0.054 & 13.82 &  3.60 & 180.1 &   7.5 & LX-200\\
 55604.8940 & 16.987 & 0.012 & 11.84 &  1.31 & 146.8 &   3.2 & Perkins\\
 55605.8860 & 16.869 & 0.010 & 12.05 &  0.02 & 148.8 &   0.0 & Perkins\\
 55607.3730 & 16.980 & 0.026 & 17.48 &  2.05 & 171.8 &   3.4 & AZT-8+ST7\\
 55608.2500 & 16.901 & 0.069 &  9.01 &  3.86 & 218.1 &  12.3 & LX-200\\
 55609.3790 & 16.904 & 0.107 & 22.39 &  6.46 & 133.0 &   8.3 & LX-200\\

\enddata

\tablecomments{ RJD=JD-2400000.0; Table \ref{tbl-1}is published in its entirety in the electronic edition of the Journal. A portion is shown here for guidance regarding its form and content.}

\end{deluxetable*}

Unlike all of the previous years, starting from 
the end of February 2011 a smooth rotation of $\chi$ (Fig.~\ref{fig2}, bottom panel) with an amplitude of $\sim330\degr$
is prominent. We see a steady rotation of $\chi$ $\sim13.3\degr$ per day during March 2011. The rotation stops at RJD~55643 (2011 March 22), near the peak of the $R$-band outburst. After that, only minor changes of EVPA are observed, despite continued strong variability of the flux density and fractional polarization. After RJD$\sim55660$ the EVPA rotates back to a ``quiescent'' state ($\sim0\degr$).

During two nights, on March 9 and April 24, 
we observed violent intranight variability, $\sim0\fm7$ within 7 hours and $\sim1\fm0$ within 5 hours, respectively, accompanied by synchronous 
changes in the fractional polarization (marked by magnified symbols in Fig.~\ref{fig2}). The fractional polarization varied from 5.8\% to 12.6\% on March 9 and from 19.8\% to 28.9\% on April 24. These are the fastest flux and polarization changes recorded for this source in the published literature.
 
\begin{figure}[ht]
\begin{center}
   \includegraphics[width=8cm,clip]{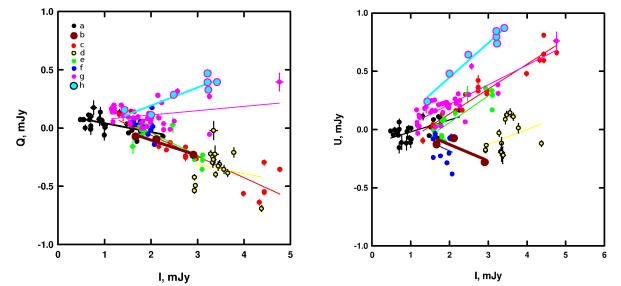}
        \caption{Absolute Stokes parameters variation during 2011 January-April; {\it left:}
        Stokes Q  vs. I, {\it right:} Stokes U vs. I. Different colors refer to 
        different stages of the evolution in ($I, Q, U$) parameter space (see Table \ref{tbl-2}). \label{fig3}}
\end{center} 
\end{figure}

Following \citet{1999BaltA...8..575H}, we plotted ($Q$ vs $I$) and ($U$ vs $I$) Stokes polarization parameters 
(see Fig.~\ref{fig3}) and found that the entire data set can be split into sections with its own behavior 
in ($I, Q, U$) parameter space. We mark these sections with different colors in Figure~\ref{fig3} and apply 
the same colors to the data plotted in Figure~\ref{fig2}.

The regression lines in Figure~\ref{fig3} represent components, each with constant parameters of polarization, $P_{\rm comp}$ and $\chi_{\rm comp}$, while its total and polarized fluxes vary. There are
8 different components with respect to the Stokes parameters behavior. Since these components are variable in flux,
we will refer to them as variable sources. We notice that the regression lines tend to converge on the locus of points corresponding 
to the pre-outburst values of the Stokes parameters. This implies that one of the components, probably 
responsible for the flux and polarization of S4\, 0954+658 before the outburst,
has constant Stokes parameters. We estimate the constant source's parameters as R=17.8 (corresponding to flux density of 0.308~mJy after correction for interstellar extinction),
p=15\% and $\chi= -6\degr$. We assume that the component should contribute the same amount of the total and polarized flux during the outburst as well. Hence, we subtract its contribution from the Stokes parameters of S4\, 0954+685 to get the radiation parameters of the variable sources. These are listed in Table~\ref{tbl-2}.

\begin{deluxetable}{cccccc}
\tablecaption{Optical Polarization Parameters of the Variable Sources \label{tbl-2}}
\tablewidth{0pt}
\tablehead{
\colhead{name}&\colhead{RJD}& \colhead{p \%}& \colhead{$\sigma$p \%}  & \colhead{$\chi$ $\degr$ } & \colhead{$\sigma \chi$ $\degr$ }
}
\startdata
a&55572-55629	& 12.66& 	1.91& 	-26.8& 	4.2\\
b&55630	        & 19.24& 	5.18& 	23.3& 	7.4\\
c&55631-55637	& 27.79& 	1.22& 	-24.5& 	1.2\\
d&55638-55639 	& 15.26&	9.00&	-24.7&	17.0\\
e&55641-55647	& 30.00& 	2.83& 	31.3& 	2.6\\
f&55651-55657	& 17.35& 	11.22& 	42.3& 	17.8\\
g&55660-55701	& 18.08& 	1.24& 	31.2& 	1.9 \\
h&55676  	& 33.05& 	2.14& 	17.0& 1.8 \\
\enddata
\end{deluxetable}

\begin{deluxetable}{cccccc}
\tablecaption{Polarization properties of knots on VLBA images\label{tbl-3}}
\tablewidth{0pt}
\tablehead{
\colhead{MJD}&\colhead{Knot}&\colhead{Flux (Jy)}& \colhead{p \%}&\colhead{$\chi$ $\degr$ }& \colhead{Date}
}
\startdata
 55724.5 & K8 &  0.10 &  27.4 & 146.2 & 12  JUN  2011\\
 55763.5 & \dots &  0.12 &  19.0 & 135.0 & 21  JUL  2011\\
 55796.5 & \dots &  0.15 &  18.7 & 116.8 & 23  AUG  2011\\
 55820.5 & \dots &  0.08 &  23.2 & 117.3 & 16  SEP  2011\\
 55850.5 & \dots &  0.08 &  23.8 & 122.6 & 16  OCT  2011\\
 55897.5 & \dots &  0.11 &  15.4 & 144.8 & 02  DEC  2011\\
\enddata
\tablecomments{$MJD=JD-2400000.5$; Table \ref{tbl-3} is published in its entirety (with parameters for all of the components) in the electronic edition of the Journal. A portion is shown here for guidance regarding its form and content.}

\end{deluxetable}

We use the technique developed by Hagen-Thorn \citep[see, e.g.,][and references therein]{2008ApJ...672...40H} to analyze the color variability of S4~0954+65.   If the variability is caused only by the flux variation but the relative spectral energy distribution (SED) remains unchanged,
then in {\it n}-dimensional flux space
$\{F_1, ..., F_n\}$ ({\it n} is the number of spectral bands used in multicolor observations) the observational points must lie on straight lines. The slopes of these lines are the flux ratios for
different pairs of bands as determined by the SED.
With some limitations, the opposite is also true: a linear relation
between observed fluxes at two different wavelengths during some period of flux variability
implies that the slope (flux ratio) does not change. Such a relation for several bands would indicate that
the relative SED of the variable source remains steady and can be derived from the slopes
of the lines.

We use magnitude-to-flux calibration constants for optical $BVRI$ bands from \citet{Mead1990}. Galactic absorption in the direction of S4~0954+65 is calculated according to Cardelli's extinction law \citep{Cardelli1989} and $A_V=0\fm38$ \citep{Schlegel1998}.

\begin{figure}[ht]
\begin{center}
    \includegraphics[width=8cm,clip]{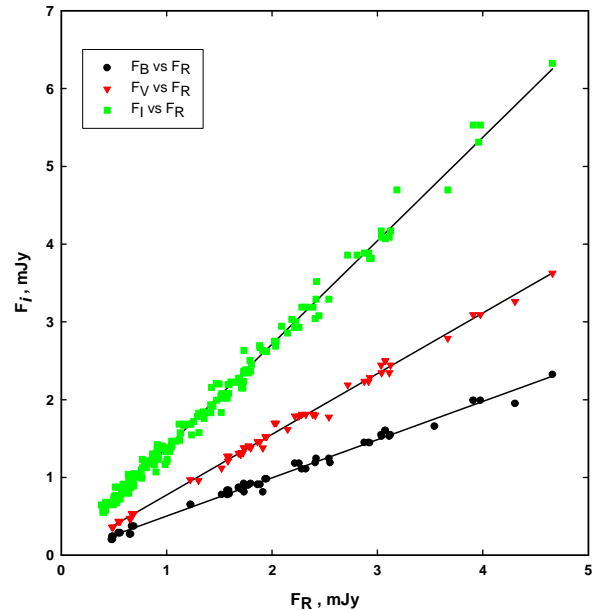}
         \caption{Dependences of the flux in B, V, and I bands on the flux in R band (the fluxes are
corrected for the Galactic extinction). The lines represent linear regression fits to the dependences. 
\label{fig4}}
\end{center} 
\end{figure}

\begin{figure}[ht]
\begin{center}
   \includegraphics[width=8cm,clip]{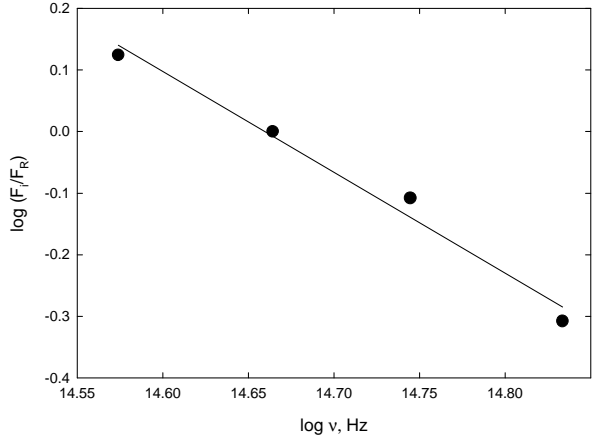}
        \caption{Relative spectral energy distribution of the variable source in S4~0954+64 obtained by using the linear regressions shown in Fig.\ref{fig4}. The solid line represents a linear fit of the 
SED. \label{fig5}}
\end{center} 
\end{figure}

Figure~\ref{fig4} presents flux-flux dependences between values in BVRI bands with R band chosen
as the primary reference band. Figure~\ref{fig4} shows that during the March-April 2011 flare the flux ratios follow linear dependences, $F_i=A_i+B_i\cdot F_R$, where i corresponds to B, V, and I bands. Values of $B_i$, the slopes of the regressions, vs. the frequency of the corresponding band represent a relative SED of the variable source. As can be seen in Fig.~\ref{fig5}, on a logarithmic scale the SED is fit very well by a linear slope $\alpha=-1.64\pm 0.15$ that suggests that the variable source emits synchrotron radiation with $F_\nu \propto \nu^{\alpha}$.

\subsection{Radio VLBI Versus Optical and Gamma-ray Data}

\begin{figure*}
\figurenum{6}
\plotone{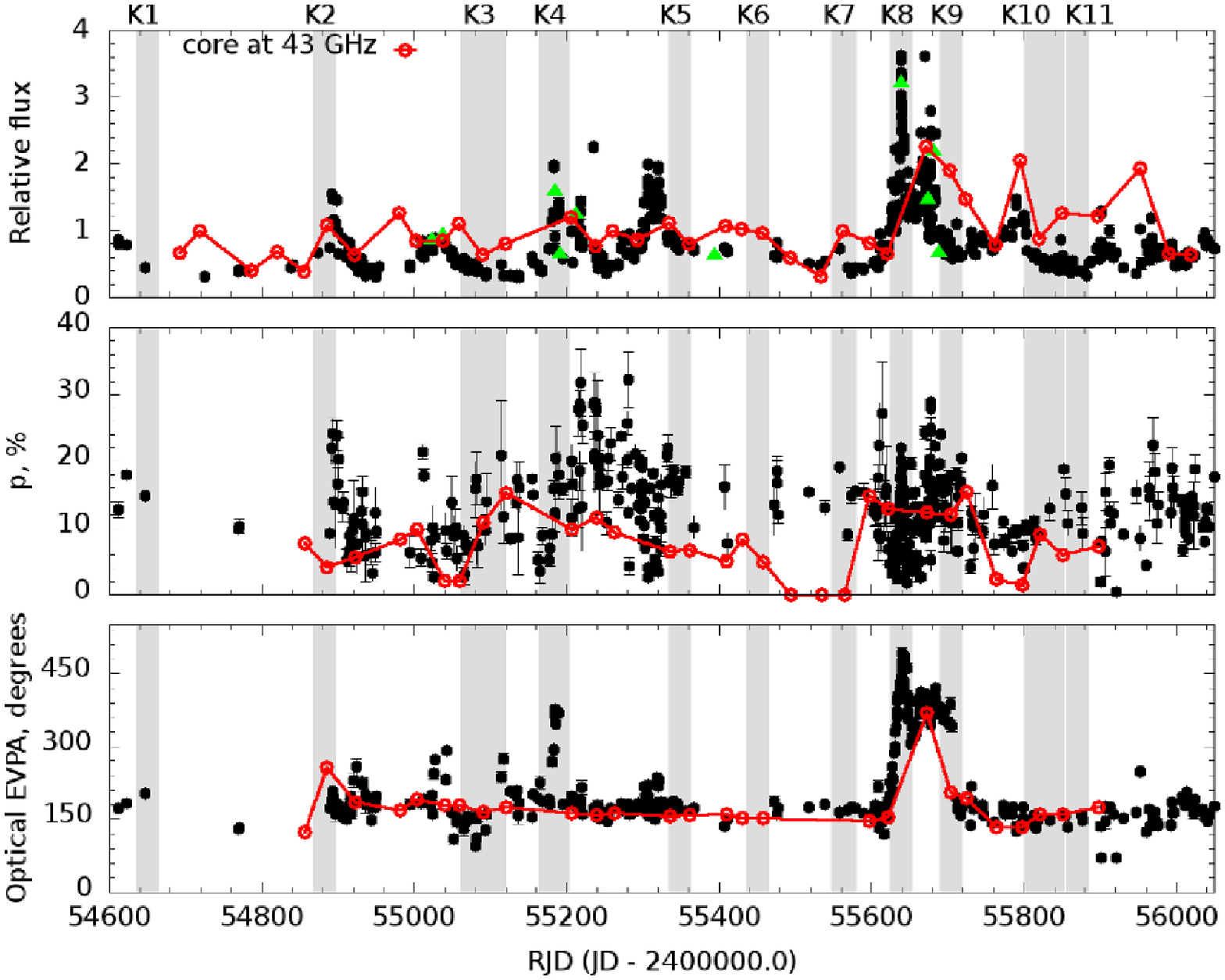}
\caption{Top panel: optical (R band) light curve (filled circles) overlaid by $\gamma$-ray light curve (triangles), and VLBI core light curve at 43 GHz (open circles). Middle panel: optical fractional polarization vs. time curve (filled circles) overlaid by $P$ of the VLBI core vs. time curve (open circles). Bottom panel: position angle of optical polarization vs. time curve overlaid by EVPA of the VLBI core vs. time curve (open circles). Plots for other components (Figs.\,6.1-6.7) are available in the electronic edition of {\it The Astronomical Journal}. \label{fig6} }
\end{figure*}

\begin{figure}[ht]
\figurenum{7}
\begin{center}
   \includegraphics[width=8cm,clip]{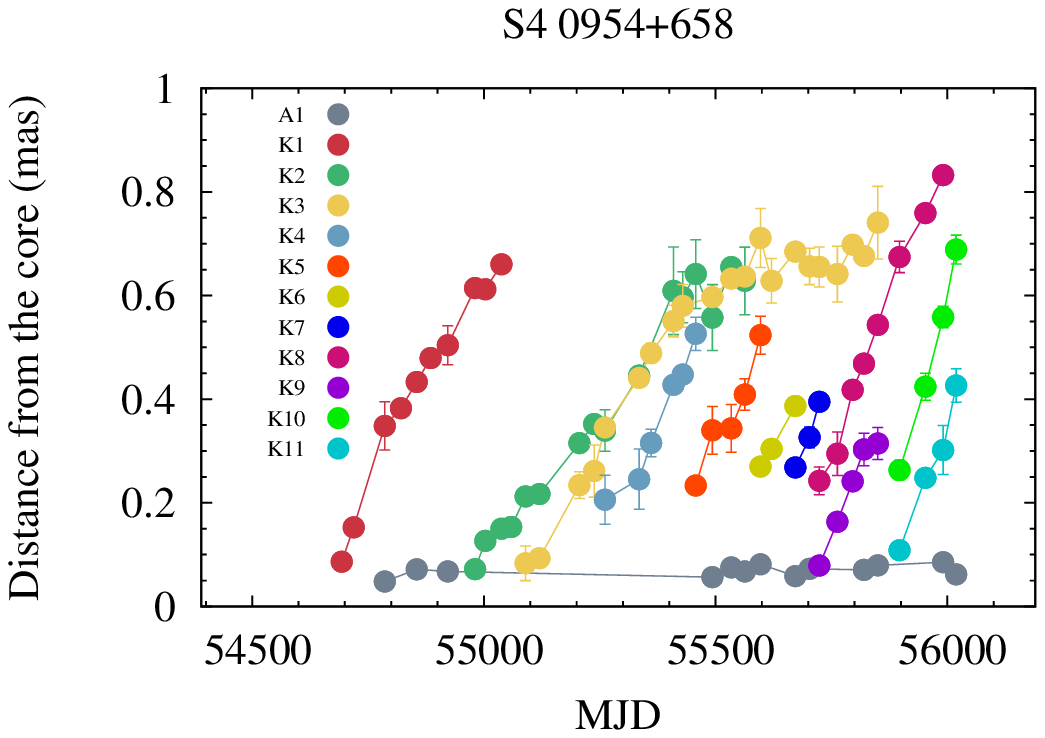}
        \caption{Separations of knots from the core as a function of time.\label{fig7}}
\end{center} 
\end{figure}

\begin{figure}[ht]
\figurenum{8}
\begin{center}
   \includegraphics[width=8cm,clip]{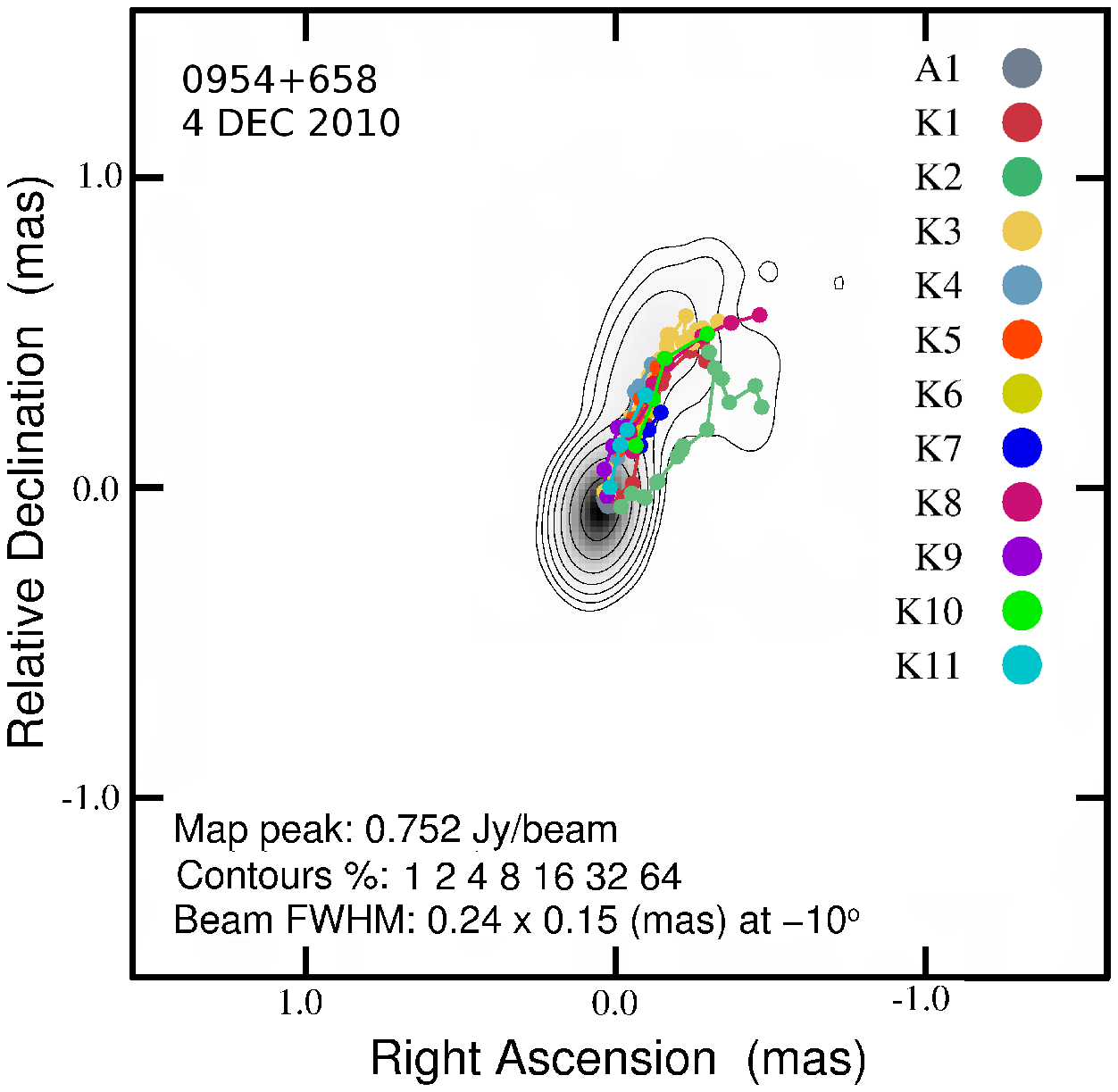}
        \caption{The 43~GHz image of the source with trajectories of knots superposed.\label{fig8}}
\end{center} 
\end{figure}

Figure~\ref{fig1} presents the multi-frequency light curves of S4\,0954+658 and optical polarization parameter 
curves along with an indication of times of ejection of the superluminal knots.  Figure~\ref{fig6} shows  the $\gamma$-ray light curve overlaid by the optical light curve (top panel); the degree of optical polarization and polarization of VLBI core at 43 GHz  (middle panel); the position angle of optical polarization and the position angle of VLBI core at 43 GHz  (bottom panel). Similar plots that show light curves and polarization parameters' curves of other VLBI knots are available on-line in the electronic edition. 
Figure \ref{fig7} shows the evolution of the distance of knots from the core, 
while Figure \ref{fig8} displays the VLBA image of the source at 43~GHz with trajectories of the knots superposed. 
   
We carefully study the optical polarization behavior of S4\,0954+658 near the ejection times of the components. 
For the majority of knots (8 of 11) we have found a connection between the time of the ejection of a component and activity at the optical and radio wavelengths (37 GHz). A visual inspection of Figure~\ref{fig6} reveals that during most of the observational period the optical EVPA was  $\chi\sim-7\degr$, close to the mean radio EVPA of the radio core ($-12\degr$) and mean jet direction ($-20\degr$).

A number of flares are apparent in the optical light curve during the period of observations 54800-56000 (Fig.\ref{fig1} ). Of particular
interest are the flares 2, 3, 3a, 5, 5a, during which  $\gamma$-ray detections occurred. To compare epochs of optical flares with the epochs of ejections of superluminal knots, we separate the sample of optical flares into 2 groups. Group A includes positive detections, for which $|(T_{\mathrm{opt_{max}}} - T_{\mathrm{eject}})|\le\sigma$, where $\sigma$ is the $1~\sigma$ uncertainty in $T_{\mathrm{eject}}$ and Group B, for which  $|(T_{\mathrm{opt_{max}}} - T_{\mathrm{eject}})|\le3~\sigma$.
Table~\ref{tbl-5} lists the epochs of optical flares (Fig.~\ref{fig1}), epochs of $\gamma$-ray detections, presence of optical $\chi$ rotation during each flare, speed of optical $\chi$ rotation if rotation is found, epoch of knot ejection if detected, and type of the flare according to classification introduced above. 

\begin{figure*}
\figurenum{9}
\begin{center}
   \includegraphics[width=18cm,clip]{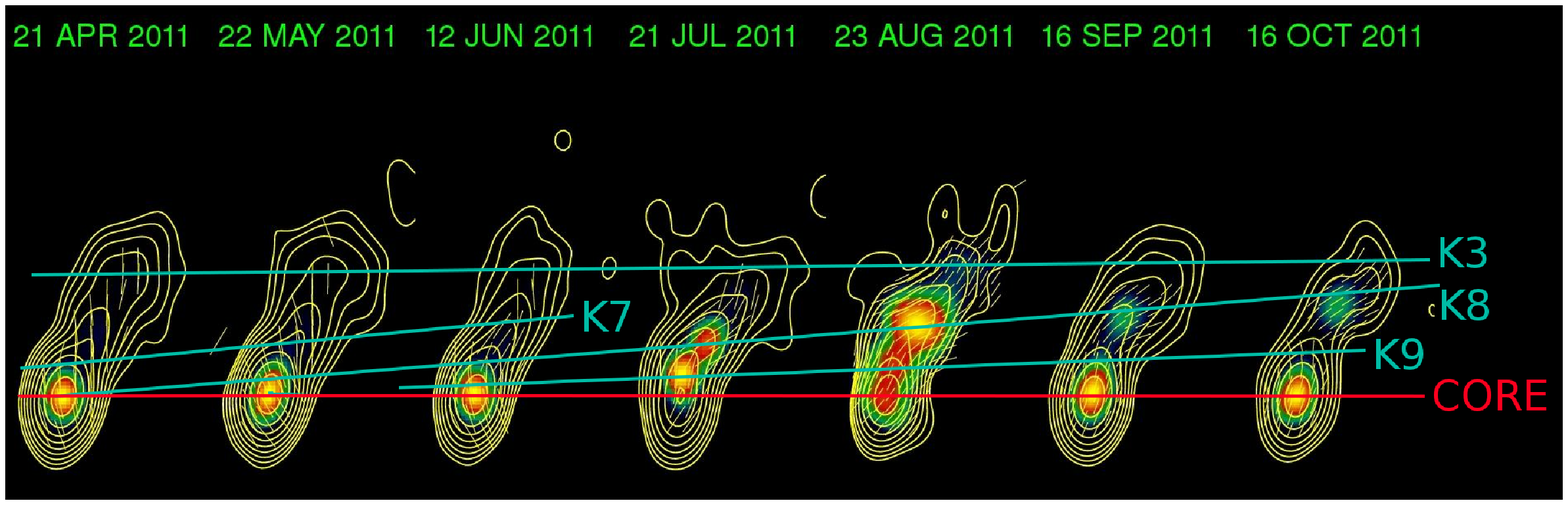}
        \caption{Total (yellow contours) and polarized (color scale) intensity images at 43~GHz; yellow line segments over the color scale show the direction of the electric vector.\label{fig9}}
\end{center} 
\end{figure*}

{\it Component K1:} Knot K1 is very bright, but we do not have enough data at optical wavelengths for a detailed analysis. Nevertheless, the 37~GHz light curve shows a strong flare that precedes the ejection time of knot K1 within $1~\sigma$ uncertainty of $T_{\mathrm{eject}}$. 

{\it Component K2:} The ejection of knot K2 was simultaneous with an optical flare and 
an increase of the optical polarization up to $24\%$ within $1~\sigma$ uncertainty of $T_{\mathrm{eject}}$. 
Although there are a number of short rotations of the optical EVPA within $3~\sigma$ uncertainty of $T_{\mathrm{eject}}$ of K2, we have too few measurements ($\le$ 4 points) to follow the EVPA evolution well in these cases. 

In addition, the position angle of K2 ($<\Theta>=-49\degr$) is quite different from the mean jet direction ($\sim -20\degr$). Before the ejection of K2 we see a modest flare in the core (RJD=54885), which coincides with the optical flare \#1 (see Table \ref{tbl-5}).  During the flare  EVPA of the core is $\sim76\degr$ that differs significantly from both the mean optical EVPA and mean  EVPA of the core ($\sim -12\degr$). There is a sharp jump in the optical EVPA at  RJD$\sim$54923 with $\chi$ varying from $78\degr$ to $45\degr$. The latter agrees with the EVPA of  K2 ($\chi=45\degr$) at RJD=54981, when the knot is first resolved from the core at the VLBA images. This suggests a connection between the optical and radio  events.

{\it Component K3:} The appearance of knot K3 was accompanied by a $\sim 27\degr$ rotation of the optical EVPA 
(RJD 55063---55068, $\sim$ 4.5 degrees/day) within $1~\sigma$ uncertainty of $T_{\mathrm{eject}}$. In addition, 
a  broad flare in R band with maximum at RJD 55024 was contemporaneous with the ejection of K3
within $3~\sigma$ uncertainty of $T_{\mathrm{eject}}$, as well as with two detections in $\gamma$-rays.

{\it Component K4:} The ejection of K4 was accompanied (within $1~\sigma$ of $T_{\mathrm{eject}}$) 
by a $\sim180\degr$ rotation of the optical EVPA ($\sim$ 15.7 degrees/day), an increase of fractional polarization 
up to $20\%$, an optical flare, a flare in VLBI core and at 37~GHz (RJD 55192  S=$2.17\pm0.14$ Jy), and three detections 
at $\gamma$-ray energies. Also, the historical maximum level of optical fractional polarization 
(RJD=55217, P=41\%) was achieved within $2~\sigma$ uncertainty of $T_{\mathrm{eject}}$. 

{\it Component K5:} The ejection of K5 was contemporaneous with an optical flare at RJD 55319 
($S_R= 1.96$ mJy, $P=12\%$). At the time when K5 was emerging from the core we did not find 
significant smooth rotation of the EVPA, but we detected an increase of the optical fractional polarization 
in the form of a plateau with a mean value of $\sim17\%$, and a strong flare at 37~GHz.  

{\it Components K6 and K7:} Knots K6 and K7 are weak and were detected only at 3 epochs. However, they are seen clearly in the polarization maps 
(see set of Fig.~\ref{fig9}). 
We have not found contemporaneous violent activities in optical and $\gamma$-ray bands, which can be associated with these components similar to those of K2-K5.

{\it Component K8:} The most interesting is knot K8, whose appearance coincides within $1~\sigma$ uncertainty with the major flare
in the R-band light curve, a flare at $\gamma$-ray energies,   a strong flare in VLBI core and at~37 GHz. The emergence of knot K8 from the core was also accompanied by a significant rotation of the optical EVPA ($\sim330^{\circ}$, $\sim$ 13.3 degrees/day), and 
by a high level of optical fractional polarization, up to $22\%$. 

{\it Component K9:} Violent intranight variability, observed during the night of 2011 April 24 
(brightening by $\sim$0.7 mag within ~ 7 hours) was contemporaneous with the ejection of knot K9 
within $2~\sigma$ uncertainty of $T_{\mathrm{eject}}$. During this flare the flux in R band increased up to 
$2.47$ mJy and the degree of optical polarization rose up to $28\%$.

{\it Component K10:} Knot K10 was ejected after a flare in the R band light curve at RJD=55789 (within $2~\sigma$ uncertainty of $T_{\mathrm{eject}}$), which was contemporaneous also with a flare at 37~GHz (RJD =55786, $S=1.66$ Jy) and a flare in VLBI core, while
a moderate degree of both optical ($\sim10\%$) and  VLBI core polrization ($\sim2\%$)  was observed during the flare. 

{\it Component K11:} The knot K11 passed through the core within $2~\sigma$ uncertainty of $T_{\mathrm{eject}}$ before  a flare in the R band light curve at RJD=55900. We have not found contemporaneous violent activities in optical and $\gamma$-ray bands,

\begin{deluxetable*}{cccccccc}
\tablecaption{Kinematic parameters of the VLBI knots\label{tbl-4}}
\tablewidth{0pt}
\tablehead{
\colhead{Knot}& \colhead{N} & \colhead{$\mu$} & \colhead{$\beta_{app}$} & \colhead{$T_{\mathrm{eject}}$} & \colhead{$\dot{\mu_{\perp}}$} &
\colhead{$\dot{\mu_{||}}$} & \colhead{$<\Theta>$} 
}
\startdata
K1& 10 &$    0.59 \pm    0.01 $ &$   13.02 \pm    0.30 $ &$ 54650.0 \pm   15 $ &$   -0.44 \pm    0.02 $ &$   -0.78 \pm    0.03 $ &$  -26.9 \pm    5.47$\\
K2& 16 &$    0.37\pm    0.01 $ &$    8.24 \pm    0.02 $ &$ 54883.5 \pm    15 $ &- &- &$ -49.0 \pm    7.7 $\\
K3& 21 &$    0.32 \pm    0.02 $ &$    6.99 \pm    0.42 $ &$55091.6 \pm   30 $ &$   -0.18 \pm    0.01 $ &$   -0.40 \pm    0.01 $ &$  -20.2 \pm    6.6$\\
K4& 6 &$    0.61 \pm    0.06 $ &$   13.53 \pm    1.42 $ &$55184.7 \pm   20.6 $ &- &- &$  -14.3 \pm    1.5  $\\
K5& 5 &$    0.69 \pm    0.05 $ &$   15.14 \pm    1.12 $ &$55349.5 \pm   14.1 $ &- &- &$  -16.7 \pm    2.3$\\
K6& 3 &$    0.58 \pm    0.01 $ &$   12.75 \pm    0.17 $ &$55450.4 \pm    15 $ &- &- &$  -21.7 \pm    1.4 $\\
K7& 3 &$    0.87 \pm    0.06 $ &$   19.24 \pm    1.31 $ &$55564.4 \pm   16.9 $ &- &- &$  -27.2 \pm    0.68 $\\
K8& 8 &$    0.86 \pm    0.01 $ &$   18.95 \pm    0.28 $ &$55639.1 \pm    15 $ &$   -1.69 \pm    0.06 $ &$   -0.23 \pm    0.06 $ &$  -25.4 \pm    6.4 $\\
K9& 5 &$    0.78 \pm    0.06$  &$   17.22 \pm    1.39 $ &$55704.5 \pm    15 $ &- &- &$  -8.4 \pm    4.4 $\\
K10& 4 &$    1.20 \pm    0.07 $ &$   26.61 \pm    1.58$ &$55827.2 \pm   26.8$ &- &- &$  -24.1 \pm    3.1 $\\
K11& 4 &$    0.92 \pm    0.04 $ &$   20.19 \pm    0.91$ &$55871.9 \pm    15 $ &- &- &$  -14.6 \pm    2.6 $\\
\enddata
\end{deluxetable*}

\begin{deluxetable*}{cccccccc}
\tablecaption{The summary of optical flares\label{tbl-5}}
\tablewidth{0pt}
\tablehead{
\colhead{N}& \colhead{Optical flare} & \colhead{$\gamma$-ray} & \colhead{Optical $\chi$ } & \colhead{Speed of optical $\chi$ } &
\colhead{Knot ejection}& \colhead{Type} & \colhead{Connection flare - knot}
\\ 
\colhead{} & \colhead{RJD} & \colhead{} & \colhead{\textquotedblleft{rotation}\textquotedblright ($\degr$ )} & \colhead{ \textquotedblleft{rotation}\textquotedblright ($\degr$/day)} & \colhead{} & \colhead{} & \colhead{}
}
\startdata
1  &54891.807 & - &  -  &   -  & K2 &A&?\\
2  &55020.307 & Y & 27  & 4.5  & K3 &B&YES\\
3  &55182.447 & Y & 180 & 15.7 & K4 &A&YES\\
3a &55217.384 & Y &  -  &   -  & K4 &B&?\\
4  &55319.363 & - &  -  &   -  & K5 &B&?\\
5  &55637.580 & Y & 333 & 13.3 & K8 &A&YES\\
5a  &55669.434 & Y &  -  &   -  & K9 &B&?\\
6  &55789.258 & - &  -  &   -  & K10&B&?\\
7  &55900.574 & - &  -  &   -  & K11&B&?\\
\enddata
\end{deluxetable*}

The feature A1 is detected at many epochs during our VLBI observations at a stable position of $0.07\pm0.01$~mas with respect to the core (see Fig.~\ref{fig7}).
\citet{2001ApJS..134..181J} found that \textquotedblleft{stationary hot spots}\textquotedblright are a common characteristic of compact jets, 
with the majority of such features located within a range of projected distances of 1-3 pc from the core. 
These authors proposed three categories of models for stationary components in supersonic jets: 
a) standing recollimation shocks caused by imbalances between the pressure internal and external to the jet; 
b) sites of maximum Doppler beaming where a bent jet points most closely to the line of sight; and 
c) stationary oblique shocks, where the jet bends abruptly. We consider that knot $A1$ falls most likely 
in the category $a$, since it is quasi-stationary with an observed \textquotedblleft{lifetime}\textquotedblright at least several months.

\subsection{Statistical analysis of coincidences between optical flares and ejections of VLBI knots}

\begin{figure}[ht]
\figurenum{10}
\begin{center}
   \includegraphics[width=8cm,clip]{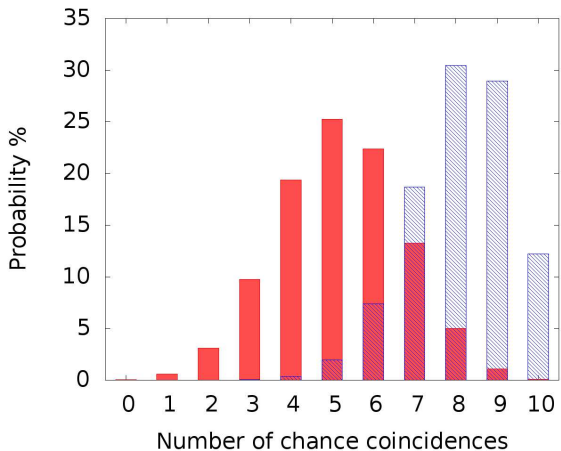}
        \caption{Probability  of chance coincidences between optical flares and epochs of zero separation within 1$\sigma$ (dark shading) and 3$\sigma$ (light shading) uncertainties.\label{fig10}}
\end{center} 
\end{figure}

We carried out numerical simulations in order to determine the probability of random coincidences between epochs of optical flares and ejection of superluminal knots in the same manner as described in \citet{2001ApJS..134..181J}. We fixed the number and epochs of optical flares according to Table \ref{tbl-5} and generated 1,000,000 samples of random epochs of ejections of VLBI superluminal components. Each sample consists of 10 random ejections (we do not include knot K1, which was ejected before the beginning of the optical and gamma-ray monitoring). We set the uncertainties of generated epochs of zero separations equal to the uncertainties  of observed superluminal ejections. A coincidence was registered in the same manner (groups A and B) as discussed above. In our observations we found 3 coincidences of group A and 6 of group B (see Table~\ref{tbl-5}). Figure~\ref{fig10} shows the results of the numerical  simulations, which demonstrate that the probability to have 3 or more coincidences within 1$\
sigma$ is  more than $80\%$.  The probability to have 9 or more coincidences within 3$\sigma$ (including 3 coincidences within 1$\sigma$) is $\sim40\%$. These values are too high to provide any meaningful constraints. An increase of the  probability of chance coincidences with number of ejections is caused by two factors: 1) significant number (10) of ejections during the relatively short observational interval of $\sim$1100 days (RJD 54850-55950), and 2) the sufficiently large mean value of a $3\sigma$ uncertainty of $\sim$ 57 days for one component, which corresponds to a half of the observational interval for 10 components. 

Although there is the quite high  probability that the optical flares and ejections of VLBI knots are not connected, it is essential to note that we use more than one criterion to associate optical flares with the appearance of superluminal knots. These include the relation between optical and radio polarization measurements, connection with detections of S4\,0954+658  in  $\gamma$-rays.  We consider with confidence that components K8 and perhaps K4 and K3 are associated with optical flares (5, 3, and 2, respectively) due to similarity in the optical/radio polarization behavior during the flares and structure of the $\gamma$-ray outbursts, which can be related to the structure of the inner jet. 

We can not exclude that the $\gamma$-ray flares RJD$\sim$55210 and RJD$\sim$55680 (near optical flares 3a and 5a, respectively) may still be associated with propagation of K4 and K8 down the jet. An interaction of the knots with the standing recollimation shock associated with A1 could lead to the second $\gamma$-ray flare and optical intra-night variability, similar to the case observed in the quasar 3C~454.3 \citep{2013ApJ...773..147J}. According to the proper motion, K8 should reach  A1 in 30$\pm$15~days, which is similar to the time lapse between the first and second  $\gamma$-ray flares, $\sim 42$ days. So the knot K9 may in fact be a new component generated after the interaction of K8 and A1. A similar case is observed for component K4 and $\gamma$-ray flare RJD$\sim55210$ (contemporaneous with optical flare 3a): knot K4 should reach  A1 in 41$\pm$21~days, while the time lapse between flares is $\sim 30$ days.

\section{Conclusions}\label{discuss}
The BL Lac object S4\,0954+658 has displayed very prominent optical activity starting from 2011 mid-February. 
Our photometric and polarimetric observations densely cover this period. In addition, we have 
an impressive set of VLBA images at 43~GHz that allows us to compare optical activity with the behavior of the parsec-scale jet. We conclude that:
\begin{enumerate}
\item{During the entire interval of our observations the source exhibited violent variability in optical bands and a
high level of activity in the jet at 43~GHz. We follow the ejection of new components with a rate  $\sim 3$ new knots per year. It should be noted that not many blazars show such a high frequency of ejections of superluminal knots, comparable with the scale of optical activity.}

\item{During the interval from RJD 54800-55900 we have identified  9  strong optical flares. Out of these 9 events, 4 were contemporaneous with positive detections of $\gamma$-ray emission at a flux level exceeding 5$\times$10$^{-7}$phot~cm$^{-2}$s$^{-1}$. Only one detection at $\gamma$-rays was not associated with an optical flare. }

\item{The overall behavior of the source during the most prominent optical outburst in 2011 March-April  can be explained as a superposition of radiation of a long lived component with constant Stokes parameters and a new, strongly variable one whose EVPA rotates at a rate of $\sim$13 degrees/day from the onset of the outburst until the moment of maximum flux
and then levels at $\sim 310\degr$. Corrected for $k\cdot180\degr$ ambiguity, this is equivalent to $-50^\circ$, 
which is quite different from the pre-outburst direction ($-6\degr$). This fast and monotonic rotation might 
be explained as the spiral motion of the variable source in a helical magnetic field (a new superluminal knot) 
\citep{2008Natur.452..966M,2010ApJ...710L.126M,2013ApJ...768...40L}. The VLBA images at 43~GHz show the ejection of a new, highly 
relativistic knot, K8, coincided within $1~\sigma$ uncertainty of $T_{\mathrm{eject}}$ with the major peak in the R-band light curve, a flare at $\gamma$-ray energies, and a flare in VLBI-core and at 37~GHz.}

\item{According to our optical data the polarization parameters of the variable source ($p=27\%$ $\chi=-25\degr$, \textquotedblleft{c}\textquotedblright in Table \ref{tbl-2}) are close to the polarization parameters of K8 ($p=27\%$, $\chi=-34\degr$ see Table \ref{tbl-3}) at the epoch (12 June 2011) when it was first separated from the core at the 43~GHz images (set of Figs.~\ref{fig6}). The knot preserved a high level of fractional polarization at later epochs.}

\item{According to our analysis, 8 of 11 superluminal components (K2, K3, K4, K5, K8, K9, K10, K11)  emerged during strong optical flares (within 1 to $3~\sigma$ uncertainty of $T_{\mathrm{eject}}$).
However, the Monte Carlo simulation indicates that there is no evidence from the timing of the optical flares and VLBI ejecta alone to support the claim that the two are related. We have very strong evidence to connect one superluminal component (K8) to a near-simultaneous optical flare, and some evidence of connections between at least 2 more (K4 and K3) superluminal ejecta and near-simultaneous optical flares. 
}

\item{The $\gamma$-ray outbursts, which can be associated with knots K4 and K8 based on $T_{\mathrm{eject}}$ (Fig.\ref{fig1}), reveal a double structure that might be explained by the interaction of a moving knot with the two stationary features in the inner jet, the core A0 (the first peak) and knot A1 (the second peak), which are presumably standing recollimation shocks. }

\item{ High-amplitude intranight variations were detected in both optical light and fractional polarization. 
This may reflect fine structure of the magnetic field, as would be expected, e.g. if the jet plasma is 
turbulent \citep{MAR2014}.}

\item{We have found 3 cases of smooth optical EVPA rotation that are associated with component ejections 
(see Table~\ref{tbl-5}) at high confidence supported by our well-sampled optical and VLBA data. 
The slowest rate of the optical EVPA rotation occurs during the appearance of knot K3, whose apparent speed was a factor of 2 slower than the average speed of superliminal knots in the jet. However, we cannot say that this is a common pattern without more data.}

\item{During the interval of our observations, the highest flux level of the VLBI core at 43 GHz was contemporaneous
with the major optical outburst. High level of fractional polarization ($\sim 13\%$) was seen in the core
during the optical flare  and dropped to 2\% after the outburst. 
A lower level of fractional polarization at 43~GHz with respect to the optical degree of polarization 
may be due to a larger volume of the region radiating at 43~GHz and turbulent magnetic field.
In addition, the polarization position angle of the core and almost all of the components 
was close to the mean jet direction, as was the optical EVPA in quiescent states 
(see set of Figs.~\ref{fig6}). This implies that the magnetic field in the regions 
of optical and radio emission has similar structure.  Moreover, a simultaneous increase of the degree of optical 
polarization and that of the core leads to the conclusion that the two regions are co-spatial.}
\end{enumerate}

\begin{acknowledgements}
We thank anonymous referee for his/her  useful comments and suggestions. This work was partly supported by Russian RFBR grants 12-02-00452, 12-02-31193, 13-02-00077, St.Petersburg University research grants 6.0.163.2010, 6.38.71.2012 and 
by NASA Fermi Guest Investigator grants NNX08AV65G, NNX11AQ03G, and NNX12AO90G. 
The VLBI data were obtained within the program VLBA-BU-BLAZAR.
The VLBA is an instrument of the National
Radio Astronomy Observatory. The National Radio Astronomy Observatory is 
a facility of the National Science Foundation operated under cooperative agreement by Associated Universities, Inc. The PRISM camera at Lowell Observatory
was developed by K. Janes et al. at BU and Lowell Observatory,
with funding from the National Science Foundation, BU, and Lowell Observatory.
The Liverpool Telescope is operated on
the island of La Palma by Liverpool John Moores University
in the Spanish Observatorio del Roque de los Muchachos of
the Instituto de Astrofisica de Canarias, with financial support
from the UK Science and Technology Facilities Council. This paper is partly based on observations carried out at the German-Spanish Calar Alto Observatory, which is jointly operated by the MPIA and the IAA-CSIC. Acquisition of the MAPCAT data is supported by MINECO (Spain) grant and AYA2010-14844, and by CEIC (Andaluc\'{i}a) grant P09-FQM-4784.
The Mets\"ahovi team acknowledges the support from the Academy of Finland
to their observing projects (\#\# 212656, 210338, 121148, and others)
This work was partly carried out using computer resources provided by Resource Center ``Computer Center of SPbU'' (http://cc.spbu.ru)

\end{acknowledgements}

\clearpage

\end{document}